\documentclass[aps,prd, 10pt, nofootinbib,floats,
floatfix,preprintnumbers,groupedaddress]{revtex4-1}
\usepackage{comment}
\usepackage{graphicx}
\usepackage{amsmath, mathrsfs}
\usepackage{float}
\usepackage{subfigure}
\usepackage[usenames]{color}
\usepackage{amssymb}
\usepackage{bbm}
\usepackage{csquotes}
\usepackage{color}
\usepackage{psfrag}
\usepackage{graphicx}
\usepackage{marginnote}
\usepackage[utf8]{inputenc}
\usepackage{tikz}
\usetikzlibrary{positioning,decorations.pathmorphing}

\usepackage{epstopdf}
\newcommand{\bea}{\begin{aligned}}
	\newcommand{\eea}{\end{aligned}}
\newcommand{\beq}{\begin{equation}}
	\newcommand{\eeq}{\end{equation}}

\newcommand{\bse}{\begin{subequations}}
	\newcommand{\ese}{\end{subequations}}

\usepackage{hyperref}
\hypersetup{
	colorlinks   = true,
	citecolor    = red,
	linkcolor    = blue,
	urlcolor     = blue,
}

\newcommand{\bmm}{\begin{multline}}
	\newcommand{\emm}{\end{multline}}

\begin{document}
	
	\title{ Probing phase transitions and microscopic interactions in quasi-topological black holes}
	
	\author{Apurba Tiwari}
	\email{apurba.e17565@cumail.in}
	\affiliation{Department of Physics, Chandigarh University, Mohali, Punjab 140413, India}
	
	\author{Randeep Kaur}
	\email{randeep.e13927@cumail.in}
	\affiliation{Department of Physics, Chandigarh University, Mohali, Punjab 140413, India}

	\author{Aruri Devaraju}
	\email{adr.me@kitsw.ac.in}
	\affiliation{ Kakatiya Institute of Technology and Science, Yerragattugutta, Warangal, Telangana 506015, India}
	
	\author{Jaya Prakash Kode}
	\email{kodejayaprakash@gmail.com}
	\affiliation{Vallurupalli Nageswara Rao Vignana Jyothi Institute of Engineering and Technology Hyderabad, Telangana, India}
	
	\author{Apparao Damarasingu}
	\email{apparao22@gmail.com}
	\affiliation{Aditya Institute of Technology and Management Tekkali, Andhra Pradesh 532201, India}
	
	\author{Silamanthula Hari Krishna}
	\email{sharikrishna@kluniversity.in}
	\affiliation{Koneru Lakshmaiah Education Foundation, Green Fields, Vaddeswaram, Andhra Pradesh 522302, India}
	
	\author{Akshay Gharat}
	\email{akshay.gharat333@gmail.com}
	\affiliation{Department of Space Engineering, Ajeenkya DY Patil University Pune, Maharashtra, India}
		
	\vspace{0.5cm}
	
	\begin{abstract}
		In this paper, we examine the thermodynamic geometry of four-dimensional quasi-topological black holes by computing the Ruppeiner scalar curvature $\mathcal{R}$ which serves as an empirical tool to describe the nature of interactions among black hole microstructures. In four dimensions, we write novel black hole solutions within the framework of generalized quasi-topological gravity, extended through a fundamental $p$-form field. Temperature, entropy, and thermodynamic volume are explicitly expressed using the extended first law. The nature of the interactions between the microstructure is then revealed by computing $\mathcal{R}$, where positive curvature indicates repulsion dominant interactions and negative curvature indicates the dominance of attraction. Our approach uses divergences and sign changing nature of $\mathcal{R}$ to identify critical points and phase transitions. Further, our analysis reveals a notably streamlined thermodynamic behavior, a single zero‑crossing of curvature $\mathcal{R}$, marking a second‑order phase transition and offering direct insight into the underlying microstructure interactions.
	\end{abstract}
	
	
	
	
	\maketitle
	\section{Introduction}
From theoretical ideas in classical general relativity to the essential tools used today in theoretical physics, black holes and their thermodynamics have undergone significant advances. Exploring the detailed connections between gravitation, quantum theory, and statistical mechanics has relied significantly on black holes in recent years. These objects offer rich and tractable models for probing the microscopic structure of spacetime and the nature of gravitational interactions beyond Einstein's theory, particularly in the context of higher-curvature gravity theories such as quasi-topological gravity~\cite{Cisterna:2020rkc,Figueroa:2020tya,Cano:2020qhy,Lei:2020clg,Myung:2020ctt,Cisterna:2021ckn,Li:2022vcd,Barrientos:2022uit}. The addition of higher-derivative curvature terms, inspired by effective field theories and string theory corrections, has extended the range of black hole geometries with nontrivial thermodynamic and phase behavior, even though classical solutions such as the Schwarzschild and Kerr black holes still offer key insights. Specifically, quasi-topological black holes offer a desirable setting in which higher-order curvature invariants can be introduced to the Einstein–Hilbert action while nonetheless generating second-order field equations in spherical symmetry. These black holes introduce new coupling parameters that regulate deviations from general relativity, thus extending the phenomenology of standard solutions. Furthermore, quasi-topological black holes are perfect testbeds to study the interplay between gravity, quantum field theory, and holography since they exhibit a range of behaviors, such as modified phase transitions, enriched thermodynamic geometry, and nontrivial microstructure interactions, when coupled with matter fields like gauge fields or scalar fields~\cite{Bueno:2022ewf,Sekhmani:2022lws,Sekhmani:2023ict,Ali:2023sul,Ali:2023vpa}.

In this paper, we study a generalization of four-dimensional quasi-topological black holes by incorporating an additional antisymmetric $(p-1)$-form gauge field $B_{[p-1]}$ to the regular Abelian Maxwell field $A_{[1]}$. The field strength of this gauge field is $H_{[p]} =dB_{[p-1]}$. Through this extension, a wider class of solutions becomes possible and the gravitational sector gains an extensive structure. In a variety of settings, these higher-rank fields gain physical interpretations inspired by supergravity and string theory. Examples include the 3-form in eleven-dimensional supergravity, the Ramond-Ramond forms in Type II theories, and the Kalb–Ramond 2-form in the heterotic string~\cite{Duan:2023gng,Masood:2024oej,Liu:2024wpa,al-Badawi:2024pdx,Anand:2025vfj,Ovgun:2025ctx,Mou:2023nrx,Hod:2024shg,Chen:2024kmy,Shahzad:2024pti}. We examine a purely magnetic $p$-form field configuration in our quasi-topological framework, which corresponds to the symmetries of a static, spherically symmetric black hole geometry. The nontrivial connection between the quasi-topological curvature terms and the higher-form fields plays an important role in the features of these new families of black hole solutions, especially in thermodynamic behavior and stability~\cite{Ali:2025sni,Ali:2025tjz}. Interestingly, even in four dimensions, where dyonic configurations are frequently limited, these couplings expand the solution space beyond the minimal quasi-topological setting examined in previous publications like \cite{Lovelock,Brigante:2007nu,Liu:2019rib}. The incorporation of higher-form sectors demonstrates the versatility of quasi-topological gravity in capturing generalized matter couplings, even if our focus here is still on neutral or electrically charged black holes in four dimensions. The framework for investigating the consequent Ruppeiner geometry, thermodynamic phase structure, and microstructure interactions in such changed gravity backgrounds is laid by this configuration.

A fundamental thermodynamic phenomenon in black hole physics is the Hawking-Page phase transition, which describes a transition from a thermal AdS spacetime to a stable large black hole phase. Within the framework of the AdS/CFT correspondence, this transition is widely interpreted as the dual of a confinement/deconfinement transition in the boundary gauge theory. In the context of the AdS/CFT correspondence, where quasi-topological black holes in asymptotically AdS spacetimes serve as dual representations of strongly coupled quantum field theories, the significance of such configurations is further enhanced. In addition to expanding our understanding of thermodynamic stability and black hole entropy, these solutions offer significant novel insights on quantum criticality, information transfer, and renormalization group flows in the dual theory. Consequently, quasi-topological black holes remain essential for connecting the frontiers of high-energy theory with gravitational physics.

A rich underlying microstructure, where microscopic degrees of freedom are reflected by the Bekenstein–Hawking entropy, emerges when black hole horizons acquire a temperature. In quasi‑topological gravity, these microstructures demonstrate thermal behavior akin to molecules, following equipartition principles. By embedding thermodynamic fluctuations into a Riemannian metric, thermodynamic geometry, and more especially Ruppeiner geometry, offers a strong framework to investigate such interactions. Using internal energy and electric potential as fluctuation variables, Ruppeiner geometry was originally applied to black holes~\cite{Cai:1998ep} and then generalized to charged (Reissner–Nordström) and rotating (Kerr) AdS solutions~\cite{Janyszek:1989zz,Ruppeiner:1981znl,Ruppeiner:2008kd,Ruppeiner:2010dzw,Dey:2011cs,Ruppeiner:2012ibw}. Depending on charge and coupling, these studies demonstrated that charged AdS black holes may transition from repulsion-dominated $(\mathcal{R}>0)$ to attraction-dominated $(\mathcal{R}<0)$ microstructures, and that neutral cases are typically attractive-dominated in higher-curvature models such as Gauss-Bonnet gravity~\cite{Weinhold:1975fyh,Weinhold:1975xej,Janyszek:1990wdh,Quevedo:2007mj,Sarkar:2006tg,Sarkar:2008ji,Sahay:2010tx,Bellucci:2008kh}.

The extension of the ADM mass to asymptotically AdS spacetimes is an important achievement in the thermodynamics of black holes, particularly in modified gravity models such as quasi-topological gravity. The interpretation of the cosmological constant $\Lambda$ as a dynamical thermodynamic variable represents a significant paradigm change in this extended thermodynamic concept. The work in \cite{Kubiznak:2012wp,Kastor:2009wy} served as inspiration for this reinterpretation, which elevates $\Lambda$ to the position of a pressure term in the bulk and defines it as,
\begin{equation}
	P = -\frac{\Lambda}{8\pi},
\end{equation}
where the thermodynamic volume $V$ is its conjugate variable. The Smarr relation and the first law of black hole thermodynamics should be modified in accordance with this extension, offering the phase space a richer structure~\cite{Gregory:2017sor,Mann:2025xrb}. The geometric and thermodynamic features of black hole solutions are significantly altered by higher-curvature corrections in the context of four-dimensional quasi-topological AdS black holes. Higher-order curvature invariants in the gravitational action cause these corrections, which are governed by coupling parameters like $\alpha$, the quasi-topological coupling~\cite{AyonBeato:1998ub,AyonBeato:1999ec,AyonBeato:1999rg,Babichev:2020qpr,Frolov:2016pav,Spallucci:2017aod,Feng:2015sbw,Cisterna:2020kde}. Standard thermodynamic parameters, including temperature, entropy, and specific heat, are therefore changed, resulting in new phase behavior, such as reentrant phase transitions and critical points. The thermodynamic behavior is further enhanced by adding electromagnetic fields (unless magnetic monopoles are invoked), which reveals a sensitive dependence on the quasi-topological couplings. A microscopic interpretation consistent with the Bekenstein–Hawking entropy is not only required but necessary since these quasi-topological black holes follow a generalized first law in the extended phase space and admit a well-defined Hawking temperature. Thermodynamic geometry, especially Ruppeiner geometry, offers a strong analytical framework for this study.   It offers the space of equilibrium states a Riemannian structure, with the Hessian of the entropy with respect to extended variables serving as the starting point of the line element. The Ruppeiner scalar curvature $\mathcal{R}$, which captures statistical interactions between the black hole's microstructures, may be calculated using this approach. When $\mathcal{R}$ is positive, it usually means repulsive interactions, like in Fermi gases, and when it is negative, it means attractive interactions, like in Bose systems. Similar to classical perfect gases, non-interacting components are indicated by a vanishing $\mathcal{R}$. Additionally, $\mathcal{R}$ divergences are often seen as evidence of critical phenomena or second-order phase transitions. These divergences are consistent with those of the compressibility and specific heat in quasi-topological black holes, so validating the connection between phase transitions in black hole systems and the curvature singularities in thermodynamic geometry. As a result, Ruppeiner geometry provides an excellent perspective for deeply investigating the interaction between underlying quantum gravity microphysics and macroscopic thermodynamics.

{\bf Motivations and Plan:}
Within the paradigm of extended black hole thermodynamics, we examine the thermodynamic geometry of dyonic quasi-topological black holes in $D=4$ anti-de Sitter (AdS) spacetime in this work. This framework uses the enthalpy $H$  of the spacetime \cite{Kubiznak:2012wp,Kastor:2009wy,Gregory:2017sor} to determine the mass $M$ of the black hole. This makes it a desirable thermodynamic potential for examining the microscopic structure using Ruppeiner geometry. We analyze the thermodynamic phase space defined by \((T, V)\), where \(V\) is the thermodynamic volume and \(T\) is the Hawking temperature\footnote{There can be a plethora of other fluctuation coordinates and planes such as \((T, P)\) or \((S, P)\) where one can investigate the Ruppeiner curvature but the \((S, V)\)-planes is not suitable as entropy and volume are not independent}. We can describe the nature of microscopic interactions in terms of the Ruppeiner scalar curvature \( \mathcal{R} \) by using the Ruppeiner metric, which is constructed from the Hessian of entropy with respect to extended variables.

Our work is organised as follows: In section-\ref{Quasi-review}, we review the basic properties of quasi-topological field theory in four dimensions. In section-\ref{Thermodynamics1}, we discuss the thermodynamics of four dimensional quasi-topological black holes in extended phase space. In section-\ref{Thermodynamic geometry}, we compute the thermodynamic curvature of quasi-topological black holes and investigate the nature of interactions among the black hole microstructurs. Finally, we will conclude with remarks in section-\ref{Remarks}. Throughout the paper, we set $k_B = c= \hbar = 1$.

\section{Review of quasi-topological field theory in four dimensions}\label{Quasi-review}

Building upon the framework in \cite{Liu:2019rib}, we construct a quasi-topological theory in four dimensions using the Maxwell field strength $\mathbb{F}_{[2]} = \mathrm{d}\mathbb{A}_{[1]}$ and a higher-rank 2-form field strength $\mathcal{H}_{[2]} = \mathrm{d}\mathcal{B}_{[1]}$. These give rise to the gauge-invariant scalars:
\begin{align}
	|\mathcal{F}_{(2)}|^2 &\propto \delta^{\mu_1 \mu_2}_{\nu_1 \nu_2} \mathbb{F}_{\mu_1\mu_2} \mathbb{F}^{\nu_1\nu_2}, \\
	|\mathcal{H}_{(2)}|^2 &\propto \delta^{\mu_1 \mu_2}_{\nu_1 \nu_2} \mathcal{H}_{\mu_1\mu_2} \mathcal{H}^{\nu_1\nu_2}, \\
	|\mathcal{I}_{(4)}|^2 &\propto \delta^{\mu_1 \ldots \mu_4}_{\nu_1 \ldots \nu_4} \mathbb{F}_{\mu_1\mu_2} \mathcal{H}_{\mu_3\mu_4} \mathbb{F}^{\nu_1\nu_2} \mathcal{H}^{\nu_3\nu_4}.
\end{align}
Restricting to configurations with purely electric $\mathbb{F}_{\mu\nu} \propto \phi'(r) \delta^{01}_{\mu\nu}$ and purely magnetic $\mathcal{H}_{\mu\nu} \propto \delta^{23}_{\mu\nu}$ components, the relevant action simplifies to:
\begin{align}\label{eq:4Daction}
	\mathcal{I}_4 = \int \mathrm{d}^4x \sqrt{-g} \left[ \mathcal{L}_\text{QT} - \left(\frac{1}{4} \mathbb{F}^2 + \frac{1}{8} \mathcal{H}^2 + \alpha \mathcal{L}_I \right) \right],
\end{align}
where the interaction term is
\begin{align}
	\mathcal{L}_I = \delta^{\mu_1\ldots\mu_4}_{\nu_1\ldots\nu_4} \mathbb{F}_{\mu_1\mu_2} \mathcal{H}_{\mu_3\mu_4} \mathbb{F}^{\nu_1\nu_2} \mathcal{H}^{\nu_3\nu_4},
\end{align}
and the quasi-topological Lagrangian is
\begin{align}
	\mathcal{L}_\text{QT} = R - 2\Lambda + \beta \delta^{\mu\nu\rho\sigma}_{\alpha\beta\gamma\delta} R^{\alpha\beta}_{\phantom{\alpha\beta}\mu\nu} R^{\gamma\delta}_{\phantom{\gamma\delta}\rho\sigma}.
\end{align}
The resulting field equations are:
\begin{eqnarray}
	\mathcal{G}_{\mu\nu} &=& \mathcal{E}_{\mu\nu}^{(0)} + \beta \mathcal{E}_{\mu\nu}^{(2)} - \frac{1}{2} \mathbb{F}_{\mu\rho}\mathbb{F}_{\nu}^{\ \rho} + \frac{1}{8} g_{\mu\nu} \mathbb{F}^2 - \frac{1}{4} \mathcal{T}^{(\mathcal{H})}_{\mu\nu} - \frac{\alpha}{2} g_{\mu\nu} \mathcal{L}_I, \\
	\nabla_{\nu} \mathbb{F}^{\nu\mu} &=& 4 \alpha \delta^{\mu\nu\rho\sigma}_{\lambda_1\ldots\lambda_4} \mathcal{H}_{\rho\sigma} \nabla_{\nu}( \mathbb{F}^{\lambda_1\lambda_2} \mathcal{H}^{\lambda_3\lambda_4} ), \\
	\nabla_{\mu} \mathcal{H}^{\mu\nu} &=& -4 \alpha \delta^{\mu\nu\rho\sigma}_{\lambda_1\ldots\lambda_4} \mathbb{F}_{\mu\nu} \nabla_{\rho}( \mathbb{F}^{\lambda_1\lambda_2} \mathcal{H}^{\lambda_3\lambda_4} ).
\end{eqnarray}
The energy-momentum tensor for the 2-form field is,
\begin{align}
	\mathcal{T}^{(\mathcal{H})}_{\mu\nu} = \mathcal{H}_{\mu\rho} \mathcal{H}_{\nu}^{\ \rho} - \frac{1}{4} g_{\mu\nu} \mathcal{H}^2.
\end{align}
Variation of the interaction term obeys,
\begin{align}
	\frac{1}{\sqrt{-g}} \frac{\delta (\sqrt{-g} \mathcal{L}_I)}{\delta g^{\mu\nu}} = \frac{1}{2} g_{\mu\nu} \mathcal{L}_I,
\end{align}
which simplifies the derivation of field equations. We now construct exact black hole solutions to the quasi-topological theory defined by the action \eqref{eq:4Daction} for the four-dimensional case ($D=4$). Despite its nonlinearities, this system admits analytic integration even in the presence of both electric and magnetic charges. Consider the static, spherically symmetric spacetime metric,
\begin{equation}
	ds^2 = -f(r) dt^2 + \frac{dr^2}{f(r)} + r^2 d\Omega_2^2,
\end{equation}
where $d\Omega_2^2$ represents the metric on the unit two-sphere with curvature $\gamma = +1$. The higher-rank magnetic field $\mathcal{H}_{\mu\nu}$ is taken to wrap the two-sphere geometry as,
\begin{equation}
	\mathcal{H}_{\theta\phi} = Q_m \sin\theta,
\end{equation}
where $Q_m$ is the magnetic charge. The gauge field strength tensor $F_{\mu\nu}$ is purely electric,
\begin{equation}
	F_{tr} = \phi'(r).
\end{equation}
In this setting, the modified Maxwell equation from the action \eqref{eq:4Daction} becomes,
\begin{equation}
	\phi''(r) + \frac{2}{r} \phi'(r) - \frac{64 \alpha Q_m^2 \phi'(r)}{\left(r^4 + 64\alpha Q_m^2\right)} = 0,
\end{equation}
and has the general solution,
\begin{equation}
	\phi'(r) = \frac{Q_e r^2}{r^4 + 64 \alpha Q_m^2},
\end{equation}
where $Q_e$ is the electric charge. With this, the $tt$ component of the Einstein equations becomes:
\begin{equation}
	\frac{f'(r)}{r} + \frac{f(r) - 1}{r^2} + \Lambda = \frac{Q_m^2}{2r^4} + \frac{Q_e^2}{2\left(r^4 + 64\alpha Q_m^2\right)},
\end{equation}
which integrates to yield the exact solution,
\begin{equation}\label{f(r)}
	f(r)=  1 - \frac{2M}{r} - \frac{\Lambda r^2}{3} + \frac{Q_m^2}{2r^2}+ \frac{Q_e^2}{2r^2} \, {}_2F_1\left(1, \frac{1}{2}, \frac{3}{2}, -\frac{64\alpha Q_m^2}{r^4}\right),
\end{equation}
where $_2F_1$ denotes the Gaussian hypergeometric function and $M$ is the ADM mass. The interaction between electric and magnetic fields introduces nontrivial screening effects, encoded in the hypergeometric term. The solution remains regular outside the event horizon $r_+$, defined as the largest positive root of $f(r_+) = 0$. For suitable parameters, the geometry exhibits multiple horizons or extremality. In the extremal case, the near-horizon geometry becomes $\text{AdS}_2 \times S^2$. Otherwise, for non-degenerate horizons, the solution exhibits Rindler behavior near $r_+$. This completes the characterization of static, spherically symmetric, dyonic black hole solutions in the quasi-topological theory~\cite{Wheeler:1985qd,OlivaRay,Myers-Robinson,Dehghani-Mann,Quintic,Ferraro,PablosRecursive,Banados:1992wn}. We now proceed to analyze their thermodynamic behavior.

\section{Thermodynamics of  quasi-topological black holes in four dimensions}\label{Thermodynamics1}
We now focus on the thermodynamics of the four-dimensional dyonic black hole solutions in the quasi-topological framework introduced earlier. The Hawking temperature~\cite{Wald:1999vt,Bekenstein:1973ur,Bekenstein:1974ax,Bardeen:1973gs,Hawking:1975vcx}, computed from the surface gravity at the horizon $r_+$, takes the form
\begin{equation}
	T = \frac{f'(r_+)}{4\pi} = \frac{r_+}{8\pi} \left( \frac{2\gamma}{r_+^2} - 4\Lambda - \frac{Q_m^2}{r_+^4} - \frac{Q_e^2}{r_+^4 + 32\alpha Q_m^2} \right).
\end{equation}
This demonstrates the suppression of the electric field by the nonlinear coupling with the magnetic component. At large $r$, the metric function behaves as
\begin{equation}
	f(r) = -\frac{\Lambda r^2}{3} + \gamma - \frac{M}{2\sigma_\gamma r} + \frac{Q_e^2 + Q_m^2}{4r^2}- \frac{8\alpha Q_e^2 Q_m^2}{5r^6} + \mathcal{O}\left(\frac{1}{r^8}\right),
\end{equation}
which satisfies the asymptotically AdS$_4$ Brown-Teitelboim boundary conditions. The Bekenstein-Hawking entropy follows the area law,
\begin{equation}
	S = 4\pi r_+^2 \sigma_\gamma.
\end{equation}
The electric and magnetic charges are calculated via flux integrals:
\begin{equation}
	Q_e \sim \int_{\Sigma_\infty} \ast \mathcal{F}_{[2]}, \quad Q_m \sim \int_{\Sigma_\infty} \mathcal{H}_{[2]},
\end{equation}
with $\Sigma_\infty$ being a boundary sphere at spatial infinity. The first law of black hole thermodynamics is satisfied:
\begin{equation}
	dM = T dS + \Phi_e dQ_e + \Phi_m dQ_m,
\end{equation}
where the electrostatic potentials are
\begin{align}
	\Phi_e &= \frac{Q_e \sigma_\gamma}{r_+} \, {}_2F_1\left(1, \frac{1}{4}, \frac{5}{4}, -\frac{32\alpha Q_m^2}{r_+^4}\right), \\
	\Phi_m &= \frac{Q_m^2 \sigma_\gamma}{r_+} + \frac{Q_e^2 r_+^5 \sigma_\gamma}{4(r_+^4 + 32\alpha Q_m^2)Q_m} - \frac{Q_e^2 \sigma_\gamma}{4Q_m r_+} \, {}_2F_1\left(1, \frac{1}{4}, \frac{5}{4}, -\frac{32\alpha Q_m^2}{r_+^4}\right).
\end{align}

\subsection{Extended thermodynamics and phase transition of  quasi-topological black holes in four dimensions}\label{Subsection:Extended thermodynamics and phase transition of quasi-topological black holes in four dimensions}
The framework of extended black hole thermodynamics is examined in this subsection. The cosmological constant \( \Lambda \) is interpreted as a dynamical thermodynamic pressure and the black hole mass \( M \) is identified with the spacetime enthalpy\footnote{The metric on the space of thermodynamic equilibrium states is evaluated by identifying $U$ with the black hole's mass $M$ in standard black hole thermodynamics. However, $M$ is associated with enthalpy $H$ in extended thermodynamics, therefore it is reasonable to refer to it as an enthalpy representation.}, \( H= M = U + PV \). It is possible to explicitly calculate the important thermodynamic variables for the class of four-dimensional dyonic quasi-topological AdS black holes that are defined by the metric function in eqn.~\eqref{f(r)}. The following formula provides the Hawking temperature \( T \) and thermodynamic volume \( V \) in the extended phase space explicitly as,
\begin{eqnarray}\label{Temperature}
	T= 2Pr_+ +\frac{1}{4 \pi  r_+} -\frac{r_+ Q_e^2}{16 \pi  \left(32 \alpha  Q_m^2+r_+^4\right)}-\frac{Q_m^2}{16 \pi  r_+^3}; \;\;\;\;\;\;\;\;\ V=\frac{1}{2} \pi ^2 r_+^4 \ .
\end{eqnarray}
Here, \( r_+ \) indicates the event horizon's radius, and \( Q_e \) and \( Q_m \) stand for the electric and magnetic charges, respectively. Higher-curvature contributions are shown by the nontrivial \( r_+ \) dependence in the gauge sector, while the quasi-topological coupling is represented by \( \alpha \). Eqn.~\eqref{Temperature} is analyzed analytically and explicitly in order to investigate the thermodynamic behavior, specifically the relationship of the Hawking temperature on the horizon radius. In the following sections, we will analyze the critical behavior and nontrivial extrema associated with phase transitions in the resulting profile.
\begin{figure}[h!]
	\begin{center}
		\includegraphics[scale=.65]{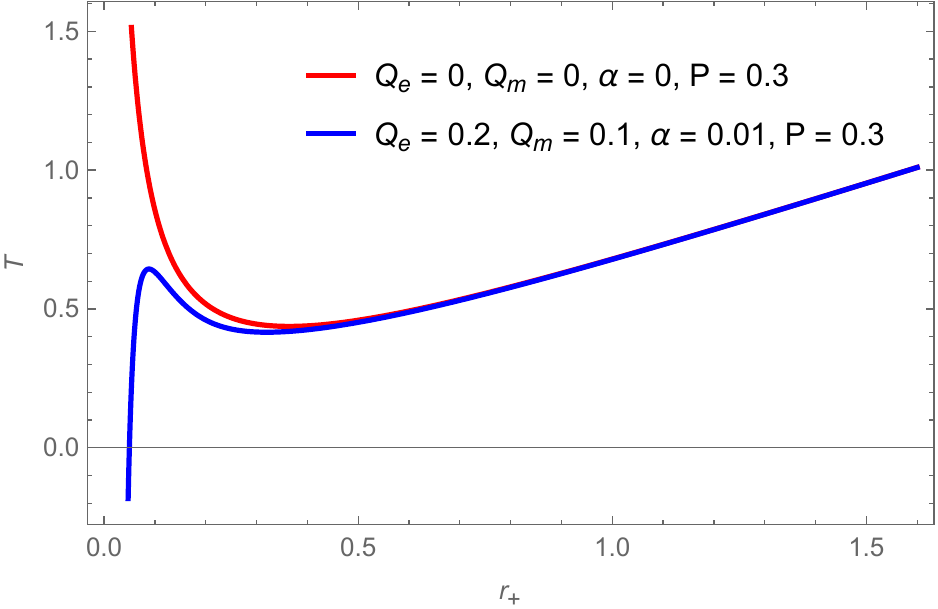} 
	\end{center}
	\caption{Behavior of temperature $T$ with horizon radius $r_{+}$ for fixed values of charges $Q_{e}, Q_{m}$, coupling $\alpha$ and pressure $P$.}
	\label{T_vs_r}
\end{figure}
For four-dimensional quasi-topological AdS black holes, the Hawking temperature $T$ plotted against the event horizon radius $r_+$ shows different thermodynamic behaviors for two configurations. The temperature increases monotonically for large $r_+$ in the Schwarzschild AdS limit, which is characterized by vanishing electric and magnetic charges ($Q_e = Q_m = 0$) and quasi-topological coupling ($\alpha = 0$). This is because the AdS pressure term, $T \sim 2P r_+$, dominates the limit. The temperature profile has a global minimum at small $r_+$ because of the $1/r_+$ term, which causes the temperature to diverge. This minimum corresponds to the well known Hawking-Page transition between thermal AdS and large black holes and suggests the existence of a critical temperature below which no black hole solutions exist. The stability structure of black holes in asymptotically AdS spacetime within Einstein gravity is reflected in this behavior, which is typical of Schwarzschild AdS black holes.

\begin{figure}[h!]
	\begin{center}
		\includegraphics[scale=.65]{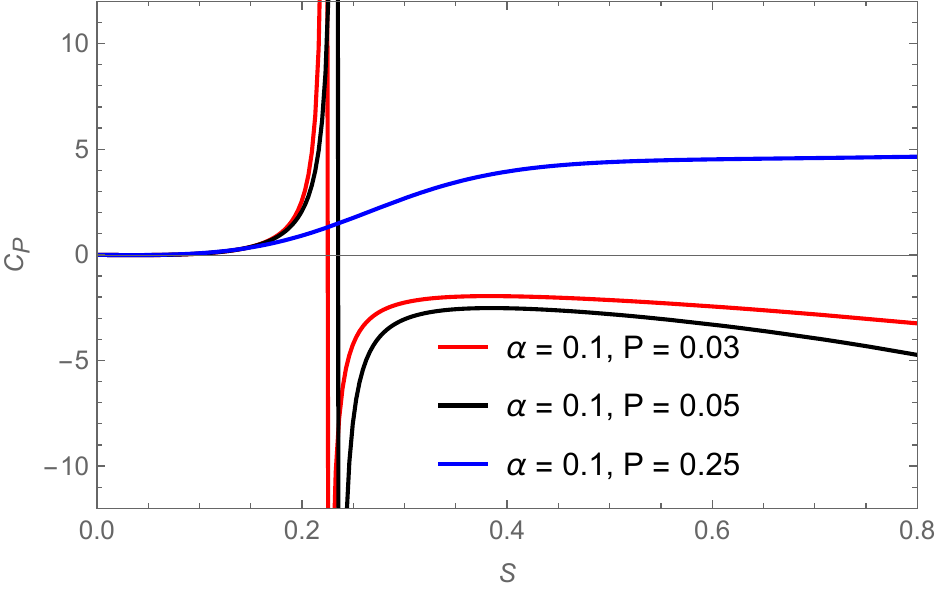} 
	\end{center}
	\caption{The behavior of the specific heat capacity $C_P$ with horizon radius $r_{+}$ for fixed values of coupling $\alpha=0.1$, charges $Q_e =0.1 , Q_m=0.3 $ and thermodynamic pressure $P$.}
	\label{C_vs_s}
\end{figure}

\begin{figure}[h!]
	\begin{center}
		\includegraphics[scale=.65]{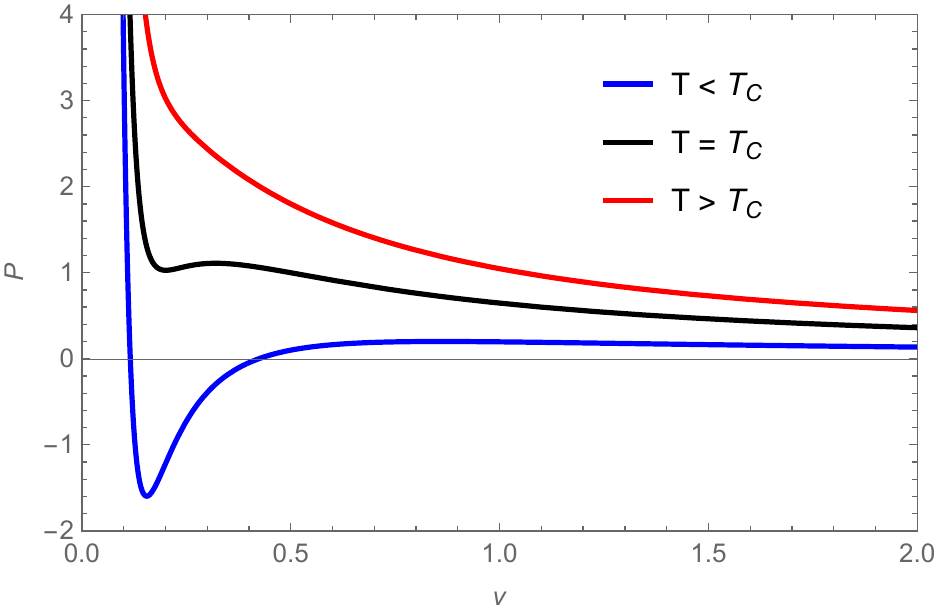} 
	\end{center}
	\caption{Behavior of pressure $P$ with specific volume $v$ for fixed charges $Q_{e}=0.1, Q_{m}=0.2$, coupling $\alpha=0.1$ and temperature $T$.}
	\label{P_vs_v}
\end{figure}

The temperature profile for the charged quasi-topological black hole case exhibits a significantly distinct structure. The effective equation of state is modified by the interaction of the electric and magnetic charges as well as the nonlinear interaction term controlled by the quasi-topological coupling $\alpha$. For a given temperature, the profile might include several branches of black hole solutions, indicating the potential for first-order phase transitions similar to the van der Waals fluid. The black hole's phase structure and microphysical interpretation are enhanced by the nonlinear interaction, which screens the electric field close to the origin and suppresses divergences. In thermodynamic geometry investigations, this behavior is especially important through Ruppeiner curvature, where the quantity and kind of zero-crossings in the scalar curvature indicate either attractive or repulsive microstructure interactions. These characteristics show how nonlinear electrodynamics and higher-order curvature affect black hole thermodynamics in the extended phase space framework. It is evident from Fig.~(\ref{T_vs_r}) that a second-order phase transition is indicated by a divergence in the black hole's specific heat capacity $C_P$ due to a minimum in the Hawking temperature. This gap indicates that black hole configurations have changed from being thermodynamically stable to unstable. The specific heat capacity $C_P$ at constant pressure $P$ can be expressed as,
\begin{equation}
	C_P = \frac{2 S \left(32 \pi ^2 \alpha  Q_m^2+S^2\right) \left(\left(4 S (8 P S+1)-\pi  Q_m^2\right) \left(32 \pi ^2 \alpha  Q_m^2+S^2\right)-\pi  S^2 Q_e^2\right)}{Q_e^2 \left(3 \pi  S^4-32 \pi ^3 \alpha  S^2 Q_m^2\right)+\left(3 \pi  Q_m^2+4 S (8 P S-1)\right) \left(32 \pi ^2 \alpha  Q_m^2+S^2\right){}^2}
	\label{Cq}
\end{equation}
The behavior of the specific heat capacity $C_P$ with the entropy $S$ of the quasi-topological black holes is plotted in Fig. \ref{C_vs_s}.
For the quasi-topological black hole, the behavior of the specific heat capacity at constant pressure, \( C_P \), as a function of entropy \( S \), provides important information about its thermodynamic stability structure. With changing thermodynamic pressure \( P \), the qualitative characteristics of \( C_P \) change considerably with a fixed value of the coupling constant \( \alpha \). In regions where \( C_P < 0 \), the specific heat exhibits non-monotonic behavior at lower pressures, suggesting local thermodynamic instability. Such regions are bounded by points where \(C_P = 0 \), which are usually connected to thresholds for phase transitions. The extent of the unstable branches decreases with increasing pressure, and for all entropies taken into consideration, the specific heat stays strictly positive at sufficiently high pressures. This behavior suggests that at higher pressures, the system achieves thermodynamic stability. A geometric view of microscopic interactions in the black hole phase space is provided by divergences in \( C_P \), which may be found close to critical points and are consistent with second-order phase transitions. These divergences frequently coincide with curvature singularities in the Ruppeiner scalar. Further, considering $\alpha$ as a variable, the first law of black hole thermodynamics\footnote{The first law for a charged black hole can be expressed as follows (in standard notation): \(dH= TdS + VdP + \Phi dQ\) where \(H= U + PV\) is the spacetime enthalpy, indicating that the entropy is a function of the parameters, i.e. \(S= S(H,Q,P)\). To keep things simple, we will just take into account the fluctuation of two thermodynamic variables and entirely fix $Q$ and $\alpha$ as mere parameters.} can be rewritten as, 
\begin{equation}
	dM = TdS + VdP + \Phi_{e} dQ_{e} +\Phi_m dQ_{m} + {\cal A}d\alpha \ ,
	\label{First law}
\end{equation}
where ${\cal A}$ is the conjugate to the coupling $\alpha$. Utilizing eqn.\eqref{Temperature} and taking the specific volume as $v=2r_+$, we can write the equation of state as,
\begin{equation}
	P=\frac{T}{v} + \frac{Q_e^2}{2 \pi  v^4 + 1024 \pi  \alpha  Q_m^2}+\frac{\left(Q_m-v\right) \left(Q_m+v\right)}{2 \pi  v^4}\ .
	\label{Eos}
\end{equation}
The behavior of thermodynamic pressure $P$ with specific volume $v$ is shown in Fig.~\ref {P_vs_v}.
In the context of extended black hole thermodynamics, the \(P-v\) diagram for quasi-topological black holes displays distinctive van der Waals-like behavior. The isotherms exhibit non-monotonic behavior for temperatures below the critical temperature \(T_c\). This is characterized by a distinctive oscillation in which the pressure \(P\) first lowers as the particular volume \(v\) increases, then increases, and then decreases once more. The existence of a first-order phase transition between small and large black hole phases is indicated by this behavior. Similar to standard fluid systems, the oscillatory region corresponds to an unstable thermodynamic phase that is usually replaced by a Maxwell equal-area structure to restore thermodynamic consistency. A second-order phase transition when the small and large black hole phases merge is indicated by an inflection point in the isotherm at the crucial temperature \(T = T_c\). The isotherms for \(T > T_c\) are monotonic and smooth, suggesting that there is no phase transition. The similarities between quasi-topological AdS black holes and van der Waals fluids are highlighted by this rich thermodynamic structure, where coupling parameters and higher curvature corrections are essential for altering the critical behavior and enhancing the phase space. Applying the condition,
\(\frac{\partial P}{\partial v}= \frac{\partial^2 P}{\partial v^2} =0,\) one can compute the inflection points\footnote{See \cite{Kubiznak:2012wp}, for detailed study of $P-V$ criticality of charged black holes in extended AdS phase space.} for a small order of $\alpha$ also. As the mass $M$ of the black hole is considered as the enthalpy, the Gibbs free energy can be computed directly as,
\begin{equation}\label{Free energy}
	F = M-TS = \frac{\left(16 \pi^2  P r_+^6 + \pi (Q_e^2+Q_m^2) +12 \pi r_+^4\right)}{r_+^2}-\frac{72 \pi  \alpha  Q_e^2 Q_m^2}{r_+^8} - \frac{32}{15} \pi ^3 r_+^3 T
\end{equation}
\begin{figure}[h!]
	\begin{center}
		\includegraphics[scale=.65]{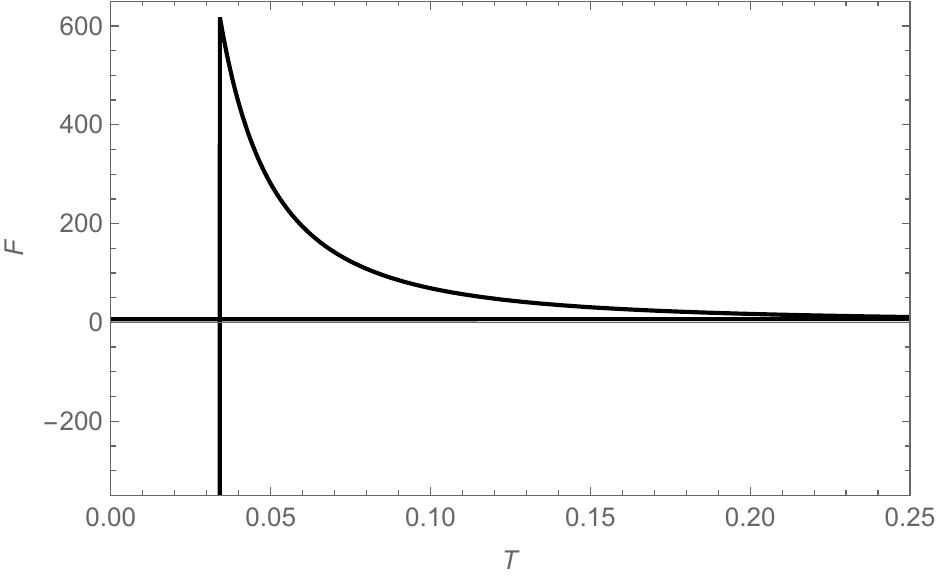} 
		
		\caption{The behavior of free energy $F$ with thermodynamic temperature $T$ for fixed values of charges $Q_{e}=0.3, Q_{m}=0.02$, coupling $\alpha=0.001$ and thermodynamic pressure $P=0.0001$.}
		\label{F_vs_T}
	\end{center} 
\end{figure}
In the context of quasi-topological black holes, the free energy \( F \) as a function of temperature \( T \), as shown in Fig.(\ref{F_vs_T}), provides important information on the system's phase structure and thermodynamic stability. The plot displays a distinctive swallow-tail structure, which is a characteristic of black hole thermodynamics first-order phase transitions. The cusp is the phase transition point where both configurations have equal free energy, and this non-analytic behavior of \( F\) illustrates the coexistence of small and large black hole phases at a critical temperature. The lower branch of the curve indicates a thermodynamically preferable small black hole phase for temperatures below this critical value, whereas the higher branch denotes an unstable or metastable large black hole phase. The system abruptly switches to the large black hole configuration as the temperature rises, and because of its lowered free energy, this configuration takes on a thermodynamical dominance. The rich phase structure brought about by higher order curvature factors and gauge interactions inherent in quasi-topological gravity is highlighted by this swallow-tail behavior, which is comparable to the liquid–gas transition in van der Waals systems. Moreover, it is consistent with the extended phase space formalism, in which the black hole mass is represented by enthalpy and the cosmological constant by pressure, proving that quasi-topological black hole thermodynamics is consistent with the well established AdS black hole physics framework.

\section{Analysis of thermodynamic geometry}\label{Thermodynamic geometry}
The thermodynamic information geometry presents an efficient method to probe the microstructure of black holes in extended phase space. This framework, established by Ruppeiner, geometrizes thermodynamic fluctuation theory and provides the manifold of equilibrium states a Riemannian metric structure~\cite{Ruppeiner:1981znl,Ruppeiner:2008kd,Ruppeiner:2010dzw,Dey:2011cs,Ruppeiner:2012ibw}. In this formalism, statistical correlations within a system can be geometrically described by constructing the metric's components from second-order derivatives of the entropy \( S \). The starting point is the Boltzmann relation:
\begin{equation}\label{eq:microstates}
	\Omega = \exp\left( \frac{S}{k_B} \right),
\end{equation}
where \( \Omega \) denotes the total number of accessible microstates and \( k_B \) is Boltzmann's constant. For a thermodynamic system characterized by two fluctuating extensive variables \( y^i \) (with \( i = 1, 2 \)), one considers a subsystem \( \mathcal{I} \subset \mathcal{I}_0 \), embedded in a larger equilibrium reservoir \( \mathcal{I}_0 \). The fluctuation probability distribution around equilibrium values is given by~\cite{RevModPhys.67.605}
\begin{equation}
	\mathbb{P}(y^1, y^2) \propto \exp\left( -\frac{1}{2} \Delta \ell^2 \right),
\end{equation}
where the line element \( \Delta \ell^2 \) defines the thermodynamic distance between nearby macrostates and is expressed as
\begin{equation}\label{eq:distance}
	\Delta \ell^2 = -\frac{1}{k_B} \frac{\partial^2 S}{\partial y^i \partial y^j} \, \delta y^i \delta y^j.
\end{equation}
The metric \( g_{ij} = -\partial_i \partial_j S \) thus encodes the Gaussian fluctuations near equilibrium and forms the foundation of the thermodynamic geometry. A smaller thermodynamic distance corresponds to a higher likelihood of fluctuations between states, while the curvature scalar \( \mathcal{R} \), derived from \( g_{ij} \), encodes the nature and intensity of microscopic interactions. Significant insights can be gained from the scalar curvature \( \mathcal{R} \), where a positive value indicates repulsive interactions and a negative value is typically regarded as reflecting attractive interactions among underlying degrees of freedom~\cite{Wei:2019yvs,Wei:2015iwa}. In systems where \( \mathcal{R} = 0 \), the ensembles do not interact. Importantly, since phase shifts are accompanied via divergent susceptibilities like specific heat or compressibility, divergences in \( \mathcal{R} \) frequently indicate these transitions. Furthermore, \( |\mathcal{R}| \sim \xi^{\tilde{d}} \) is the correlation length \( \xi \) approaching criticality, where \( \tilde{d} \) indicates the system's spatial dimensionality.

Thermodynamic geometry is particularly effective in revealing the critical behavior of electrically and magnetically charged (dyonic) black holes in the context of four-dimensional quasi-topological AdS black holes. Rich phase structure akin to classical fluids is made possible by these geometries' support for extended thermodynamic variables like electric/magnetic charges and pressure (derived from the cosmological constant). Ruppeiner curvature, as demonstrated in recent works~\cite{Wei:2019yvs,Wei:2015iwa,Ghosh:2019rsu,Ghosh:2019pwy}, reveals precise evidence for first- and second-order phase transitions, including Hawking-Page transitions and van der Waals liquid-gas transition analogues. More precisely, the behavior of \( \mathcal{R} \) for quasi-topological black holes is sensitive to the presence of both electric and magnetic charges as well as higher-curvature couplings. As a diagnostic of thermodynamic instability, the divergence of \( \mathcal{R} \) coincides with singularities in the compressibility or specific heat. Additionally, crossovers between interaction-dominated microstructures are suggested by zero-crossings or sign reversals in \( \mathcal{R} \), which match phenomena observed in Bose-Fermi gases and quantum many-body systems. These results confirm the developing consensus that thermodynamic geometry can be used to describe black hole microstructure, even though it is not immediately apparent.   In order to demonstrate how the behavior of \( \mathcal{R} \) varies over parameter space and encodes deep structural information about the underlying quantum gravitational degrees of freedom, we will use this framework to investigate dyonic solutions in quasi-topological gravity.

\subsection{Thermodynamic geometry of  quasi-topological black holes in four dimensions}\label{Sec:ThermodynamicGeometry}
A single macroscopic parameter, the horizon radius, characterizes neutral configurations such as the Schwarzschild-AdS solution in standard black hole thermodynamics. This makes the thermodynamic phase space one-dimensional and precludes a non-degenerate Riemannian geometry on the space of equilibrium states. Nonetheless, an enriched phase space is revealed under the extended thermodynamic paradigm, where the cosmological constant \( \Lambda \) is viewed as a dynamical pressure. This gives access to their underlying statistical microstructure and enables a meaningful construction of thermodynamic geometry even for neutral black holes. Following the Ruppeiner formalism~\cite{Ruppeiner:1981znl,Ruppeiner:2008kd,Ruppeiner:2010dzw,Dey:2011cs,Ruppeiner:2012ibw}, which geometrizes thermodynamics via the Hessian of entropy or enthalpy, one may define a line element on the state space in the enthalpy representation as~\cite{Wei:2019yvs,Wei:2015iwa,Ghosh:2019pwy,Ghosh:2019rsu},
\begin{equation}\label{eq:RuppeinerMetric}
d\ell^2 = \frac{1}{T} \left( \frac{\partial P}{\partial V} \right)_T \mathrm{d}V^2 + \frac{C_V}{T^2} \mathrm{d}T^2,
\end{equation}
where \( T \) is the Hawking temperature, \( V \) is the thermodynamic volume conjugate to pressure \( P \), and \( C_V \) is the specific heat at constant volume. For static, spherically symmetric black holes, \( C_V = 0 \), and hence the second term vanishes identically, rendering the thermodynamic metric degenerate. By properly rescaling the scalar curvature associated with the Ruppeiner metric, one can introduce the ``normalized thermodynamic curvature \( \mathcal{R}_N \)" in order to extract useful geometric information despite this degeneracy~\cite{Wei:2019yvs,Wei:2015iwa,Ghosh:2019pwy,Ghosh:2019rsu,Yerra:2021hnh,Singh:2023hit,Singh:2023ufh,Singh:2025ueu}.

\begin{figure}[h!]
	\begin{center}
		\includegraphics[scale=.75]{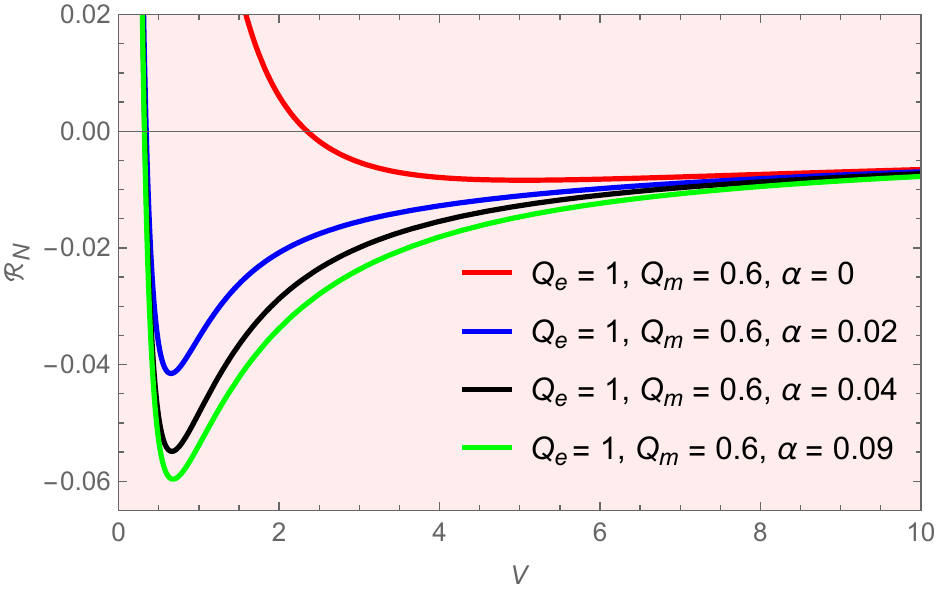} 
	\end{center}
	\caption{The behavior of thermodynamic curvature $R_N$ with volume $V$ of the black hole for fixed values of charges $Q_e, Q_m$, coupling $\alpha$ and thermodynamic temperature $T=1$.}
	\label{RN_vs_V}
\end{figure}

Further, the Ruppeiner curvature can be directly computed utilizing eqn.(\ref{Temperature}), eqn.(\ref{Eos}) and eqn.\eqref{eq:RuppeinerMetric}, and the analytical expressions of the Ruppeiner curvature for quasi-topological black holes on the $(T,V)$-plane can be expressed as,
\begin{eqnarray}\label{Curvature}
 \mathcal{R}_N = \frac{6 V Q_e^2}{\sqrt[3]{\pi } T \left(256 \pi ^{4/3} \alpha  Q_m^2+3 \sqrt[3]{6} V^{4/3}\right){}^2}+\frac{\sqrt[3]{\frac{2}{\pi }} Q_m^2}{3\ 3^{2/3} T V^{5/3}}-\frac{1}{3 \pi  T V}
\end{eqnarray}
In the limit $\alpha \rightarrow 0$ and vanishing $Q_e, Q_m$, the normalized thermodynamic curvature reduces to that of Schwarzschild case\cite{Xu:2020gud} in extended phase space given as,
\begin{eqnarray}
\mathcal{R}_N= -\frac{1}{3 \pi  T V}
\end{eqnarray}
The behavior of thermodynamic scalar curvature with volume of the black hole is shown in Fig. \ref{RN_vs_V}.
For quasi-topological black holes in AdS, the normalized Ruppeiner curvature $\mathcal{R}_N$ plotted with thermodynamic volume $V$ shows zero-crossing behavior\footnote{See \cite{May:2013yxc} for a deeper connection between zero crossings of Ruppeiner curvature and intermolecular interactions involving other potentials in various context.}. This indicates a shift in the method by which the black hole system's microstructure interactions occur. The fact that $\mathcal{R}_N$ is positive for small volumes indicates that repulsive interactions among the black hole's microscopic constituents dominate. At a specific critical volume, where repulsive and attractive interactions are balanced; an equipartition point, $\mathcal{R}_N$ falls as the volume rises and crosses zero. After this, $\mathcal{R}_N$ turns negative, suggesting that the regime of larger black hole volumes is dominated by attractive interactions. The thermodynamic richness caused by quasi-topological corrections and electromagnetic charges is highlighted by this qualitative shift from repulsion to attraction. Higher-curvature and nonlinear electromagnetic effects, which are retained in the quasi-topological coupling parameter $\alpha$, further alter the behavior here in contrast to standard Reissner–Nordström-AdS black holes, which likewise show simply one zero-crossing in $\mathcal{R}_N$. When the black hole transitions from a dense, repulsively interacting microphase to a more prolonged constructively interacting one, the degrees of freedom undergo microscopic restructuring, which is reflected in the presence of such a crossover. The location of the zero of $\mathcal{R}_N$ can be used as a diagnostic to determine criticality or the boundary between distinct interaction regimes~\cite{Zhang:2019neb,Yerra:2020oph,Wu:2020fij,NaveenaKumara:2020biu,Singh:2020tkf,Mahish:2020gwg,Anand:2025mlc}. This transition is closely linked with the system's underlying phase behavior. Thus, the following shows how thermodynamic geometry provides an accurate characterization of microstructure interactions across various black hole phases.

\section{Remarks}\label{Remarks}
In this work, we have performed a complete thermodynamic analysis of asymptotically AdS quasi-topological black holes in four dimensions that incorporate both magnetic and electric charges. We calculated and examined thermodynamic parameters including the Hawking temperature, entropy, thermodynamic volume, and specific heat within the extended thermodynamic framework, where the black hole mass is equivalent to enthalpy and the cosmological constant is viewed as a pressure. The presence of first-order small/large black hole phase transitions was confirmed by the black hole equation of state, which displayed van der Waals-like behavior on the $P$–$v$ plane with distinctive oscillations below a critical pressure. This interpretation was further supported by the free energy, which displayed a typical swallowtail structure. 

The specific heat at constant pressure, or  \( C_P \), was used to study the thermal stability of these black holes. Its behavior as a function of entropy provided important stability information. Specifically, we found that \( C_P \) diverges at critical points, indicating a phase transition of second order. \( C_P \) was negative for entropies below the divergence point, suggesting that small black holes correspond to thermodynamically unstable black hole branches. On the other hand, \( C_P \) became positive for bigger entropies, indicating stable huge black hole topologies. A distinct phase separation and stability border in the parameter space is highlighted by the division of the positive and negative heat capacity branches.

Furthermore, by examining the normalized Ruppeiner curvature \( \mathcal{R}_N \) in the extended phase space, we used the framework of thermodynamic geometry to provide insights into the microscopic interactions regulating quasi-topological black holes. When calculated in the \((T, V)\)-plane, the scalar curvature \( \mathcal{R}_N \) shows a unique behavior characterized by a single zero-crossing. The nature of dominant microscopic interactions has changed from repulsive (\( \mathcal{R}_N > 0 \)) in the small black hole branch to attractive (\( \mathcal{R}_N < 0 \)) in the large black hole regime, as indicated by this zero. A balance point where repulsive and attractive interactions cancel each other out is represented by the zero-crossing. A second-order phase transition with substantial correlation between the black hole microstructures is shown by the curvature diverging negatively near the critical volume. The relevance of quasi-topological corrections in determining the microscopic behavior is thus highlighted by Ruppeiner geometry, which not only confirms the existence of critical phenomena but also offers a geometric diagnostic of interaction types within the black hole thermodynamic ensemble.

\section*{Acknowledgements}
A.T. is grateful to Aditya Singh for useful discussions. 

\section*{Data Availability Statement}
Data sharing is not applicable to this article, as no datasets were generated or analyzed during the current study.

\section*{Conflict of Interests}
Author declare(s) no conflict of interest.


\begin{thebibliography}{99}
	
	
	\bibitem{Cisterna:2020rkc}
	A.~Cisterna, G.~Giribet, J.~Oliva and K.~Pallikaris,
	``Quasitopological electromagnetism and black holes,''
	Phys. Rev. D \textbf{101} (2020) no.12, 124041
	doi:10.1103/PhysRevD.101.124041
	[arXiv:2004.05474 [hep-th]].
	
	\bibitem{Figueroa:2020tya}
	J.~P.~Figueroa and K.~Pallikaris,
	``Quartic Horndeski, planar black holes, holographic aspects and universal bounds,''
	JHEP \textbf{09} (2020), 090
	doi:10.1007/JHEP09(2020)090
	[arXiv:2006.00967 [hep-th]].
	
	\bibitem{Cano:2020qhy}
	P.~A.~Cano and {\'A}.~Murcia,
	``Electromagnetic Quasitopological Gravities,''
	JHEP \textbf{10} (2020), 125
	doi:10.1007/JHEP10(2020)125
	[arXiv:2007.04331 [hep-th]].
	
	\bibitem{Lei:2020clg}
	Y.~Q.~Lei, X.~H.~Ge and C.~Ran,
	``Chaos of particle motion near a black hole with quasitopological electromagnetism,''
	Phys. Rev. D \textbf{104} (2021) no.4, 046020
	doi:10.1103/PhysRevD.104.046020
	[arXiv:2008.01384 [hep-th]].
	
	\bibitem{Myung:2020ctt}
	Y.~S.~Myung and D.~C.~Zou,
	``Scalarized black holes in the Einstein-Maxwell-scalar theory with a quasitopological term,''
	Phys. Rev. D \textbf{103} (2021) no.2, 024010
	doi:10.1103/PhysRevD.103.024010
	[arXiv:2011.09665 [gr-qc]].
	
	\bibitem{Cisterna:2021ckn}
	A.~Cisterna, C.~Henr{\'\i}quez-B{\'a}ez, N.~Mora and L.~Sanhueza,
	``Quasitopological electromagnetism: Reissner-Nordstr{\"o}m black strings in Einstein and Lovelock gravities,''
	Phys. Rev. D \textbf{104} (2021) no.6, 064055
	doi:10.1103/PhysRevD.104.064055
	[arXiv:2105.04239 [gr-qc]].
	
	\bibitem{Li:2022vcd}
	M.~D.~Li, H.~M.~Wang and S.~W.~Wei,
	``Triple points and novel phase transitions of dyonic AdS black holes with quasitopological electromagnetism,''
	Phys. Rev. D \textbf{105} (2022) no.10, 104048
	doi:10.1103/PhysRevD.105.104048
	[arXiv:2201.09026 [gr-qc]].
	
	
	\bibitem{Barrientos:2022uit}
	J.~Barrientos and J.~Mena,
	``Joule-Thomson expansion of AdS black holes in quasitopological electromagnetism,''
	Phys. Rev. D \textbf{106} (2022) no.4, 044064
	doi:10.1103/PhysRevD.106.044064
	[arXiv:2206.06018 [gr-qc]].
	
	\bibitem{Bueno:2022ewf}
	P.~Bueno, P.~A.~Cano, J.~Moreno and G.~van der Velde,
	``Electromagnetic generalized quasitopological gravities in (2+1) dimensions,''
	Phys. Rev. D \textbf{107} (2023) no.6, 064050
	doi:10.1103/PhysRevD.107.064050
	[arXiv:2212.00637 [gr-qc]].
	
	\bibitem{Sekhmani:2022lws}
	Y.~Sekhmani, H.~Lekbich, A.~El Boukili and M.~B.~Sedra,
	``${\textbf{D}}$-dimensional dyonic AdS black holes with quasi-topological electromagnetism in Einstein Gauss{\textendash}Bonnet gravity,''
	Eur. Phys. J. C \textbf{82} (2022) no.12, 1087
	doi:10.1140/epjc/s10052-022-11045-x
	
	\bibitem{Sekhmani:2023ict}
	Y.~Sekhmani and D.~J.~Gogoi,
	Int. J. Geom. Meth. Mod. Phys. \textbf{20} (2023) no.09, 2350160
	doi:10.1142/S0219887823501608
	[arXiv:2306.02919 [gr-qc]].
	
	\bibitem{Ali:2023vpa}
	A.~Ali,
	``Quasitopological electromagnetism, conformal scalar field and Lovelock black holes,''
	Eur. Phys. J. C \textbf{83} (2023) no.7, 564
	doi:10.1140/epjc/s10052-023-11738-x
	
	\bibitem{Ali:2023sul}
	A.~Ali and A.~{\"O}vg{\"u}n,
	``Topological dyonic black holes of massive gravity with generalized quasitopological electromagnetism,''
	Eur. Phys. J. C \textbf{84} (2024) no.4, 378
	doi:10.1140/epjc/s10052-024-12710-z
	[arXiv:2308.10742 [gr-qc]].
	
	\bibitem{Duan:2023gng}
	Z.~Q.~Duan, J.~Y.~Zhao and K.~Yang,
	``Electrically charged black holes in gravity with a background Kalb{\textendash}Ramond field,''
	Eur. Phys. J. C \textbf{84} (2024) no.8, 798
	doi:10.1140/epjc/s10052-024-13188-5
	[arXiv:2310.13555 [gr-qc]].
	
	\bibitem{Masood:2024oej}
	S.~Masood,
	``The thermodynamic profile of AdS black holes in Lorentz-violating Bumblebee and Kalb-Ramond gravity,''
	[arXiv:2411.06188 [gr-qc]].
	
	\bibitem{Liu:2024wpa}
	W.~Liu, C.~Wen and J.~Wang,
	``Lorentz violation alleviates gravitationally induced entanglement degradation,''
	JHEP \textbf{01} (2025), 184
	doi:10.1007/JHEP01(2025)184
	[arXiv:2410.21681 [gr-qc]].
	
	\bibitem{al-Badawi:2024pdx}
	A.~al-Badawi, S.~Shaymatov and I.~Sakall{\i},
	``Geodesics structure and deflection angle of electrically charged black holes in gravity with a background Kalb{\textendash}Ramond field,''
	Eur. Phys. J. C \textbf{84} (2024) no.8, 825
	doi:10.1140/epjc/s10052-024-13205-7
	[arXiv:2408.09228 [gr-qc]].
	
	\bibitem{Anand:2025vfj}
	A.~Anand, A.~Singh, A.~Mishra, S.~Noori Gashti, T.~Tangphati and P.~Channuie,
	``Black Holes in Lorentz-Violating Gravity: Thermodynamics, Geometry, and Particle Dynamics,''
	[arXiv:2507.00455 [gr-qc]].
	
	\bibitem{Ovgun:2025ctx}
	A.~{\"O}vg{\"u}n,
	``Weak gravitational lensing in Ricci-coupled Kalb{\textendash}Ramond bumblebee gravity: Global monopole and axion-plasmon medium effects,''
	Phys. Dark Univ. \textbf{48} (2025), 101905
	doi:10.1016/j.dark.2025.101905
	[arXiv:2504.07130 [gr-qc]].
	
	
	
	
	\bibitem{Mou:2023nrx}
	P.~H.~Mou, Q.~Q.~Jiang, K.~J.~He and G.~P.~Li,
	``Triple points and phase transitions of D-dimensional dyonic AdS black holes with quasitopological electromagnetism in Einstein{\textendash}Gauss{\textendash}Bonnet gravity,''
	Chin. Phys. B \textbf{33} (2024) no.6, 060401
	doi:10.1088/1674-1056/ad3342
	[arXiv:2310.08010 [gr-qc]].
	
	\bibitem{Hod:2024shg}
	S.~Hod,
	``Super-extremal black holes in the quasitopological electromagnetic field theory,''
	Eur. Phys. J. C \textbf{84} (2024) no.2, 111
	doi:10.1140/epjc/s10052-024-12454-w
	[arXiv:2404.03744 [gr-qc]].
	
	\bibitem{Chen:2024kmy}
	H.~Chen, M.~Y.~Zhang, H.~Hassanabadi, B.~C.~L{\"u}tf{\"u}o{\u{g}}lu and Z.~W.~Long,
	``Thermodynamic topology of dyonic AdS black holes with quasitopological electromagnetism in Einstein-Gauss-Bonnet gravity,''
	doi:10.1142/S0219887825502494
	[arXiv:2403.14730 [gr-qc]].
	
	\bibitem{Shahzad:2024pti}
	M.~U.~Shahzad, A.~Mehmood and A.~{\"O}vg{\"u}n,
	``Thermodynamic topological classification of D-dimensional dyonic AdS black holes with quasitopological electromagnetism in Einstein-Gauss-Bonnet gravity,''
	Eur. Phys. J. Plus \textbf{139} (2024) no.9, 806
	doi:10.1140/epjp/s13360-024-05580-7
	
	\bibitem{Ali:2025sni}
	A.~Ali and K.~Saifullah,
	``Exotic Lovelock black holes and extended quasitopological electromagnetism,''
	[arXiv:2501.15154 [gr-qc]].
	
	\bibitem{Ali:2025tjz}
	A.~Ali,
	``Black holes of quartic quasitopological gravity, conformal scalar field and extended quasitopological electromagnetism,''
	Eur. Phys. J. C \textbf{85} (2025) no.7, 764
	doi:10.1140/epjc/s10052-025-14486-2
	
	
	
	

	
	\bibitem{Lovelock}
	D.~Lovelock,
	``The Einstein tensor and its generalizations,''
	J.\ Math.\ Phys.\  {\bf 12} (1971) 498.
	doi:10.1063/1.1665613
	
	\bibitem{Brigante:2007nu} 
	M.~Brigante, H.~Liu, R.~C.~Myers, S.~Shenker and S.~Yaida,
	``Viscosity Bound Violation in Higher Derivative Gravity,''
	Phys.\ Rev.\ D {\bf 77}, 126006 (2008)
	doi:10.1103/PhysRevD.77.126006
	[arXiv:0712.0805 [hep-th]].
	
	\bibitem{Liu:2019rib} 
	H.~S.~Liu, Z.~F.~Mai, Y.~Z.~Li and H.~Lu,
	``Quasi-topological Electromagnetism: Dark Energy, Dyonic Black Holes, Stable Photon Spheres and Hidden Electromagnetic Duality,''
	arXiv:1907.10876 [hep-th].
	
	
	
	
	\bibitem{Cai:1998ep}
	R.~G.~Cai and J.~H.~Cho,
	``Thermodynamic curvature of the BTZ black hole,''
	Phys. Rev. D \textbf{60} (1999), 067502
	doi:10.1103/PhysRevD.60.067502
	[arXiv:hep-th/9803261 [hep-th]].
	
	
	\bibitem{Janyszek:1989zz}
	H.~Janyszek and R.~Mrugala,
	``Riemannian geometry and the thermodynamics of model magnetic systems,''
	Phys. Rev. A \textbf{39} (1989), 6515-6523
	doi:10.1103/PhysRevA.39.6515
	
	\bibitem{Ruppeiner:1981znl}
	G.~Ruppeiner,
	``Application of Riemannian geometry to the thermodynamics of a simple fluctuating magnetic system,''
	Phys. Rev. A \textbf{24} (1981) no.1, 488
	doi:10.1103/PhysRevA.24.488
	
	\bibitem{Ruppeiner:2008kd}
	G.~Ruppeiner,
	``Thermodynamic curvature and phase transitions in Kerr-Newman black holes,''
	Phys. Rev. D \textbf{78} (2008), 024016
	doi:10.1103/PhysRevD.78.024016
	[arXiv:0802.1326 [gr-qc]].
	
	\bibitem{Ruppeiner:2010dzw}
	G.~Ruppeiner,
	``Thermodynamic curvature measures interactions,''
	Am. J. Phys. \textbf{78} (2010) no.11, 1170-1180
	doi:10.1119/1.3459936
	[arXiv:1007.2160 [cond-mat.stat-mech]].
	
	\bibitem{Dey:2011cs}
	A.~Dey, P.~Roy and T.~Sarkar,
	``Information geometry, phase transitions, and the Widom line: Magnetic and liquid systems,''
	Physica A \textbf{392} (2013), 6341-6352
	doi:10.1016/j.physa.2013.09.017
	[arXiv:1111.6721 [cond-mat.stat-mech]].
	
	\bibitem{Ruppeiner:2012ibw}
	G.~Ruppeiner,
	``Thermodynamic curvature from the critical point to the triple point,''
	Phys. Rev. E \textbf{86} (2012) no.2, 021130
	doi:10.1103/PhysRevE.86.021130
	
	\bibitem{Weinhold:1975xej}
	F.~Weinhold,
	``Metric geometry of equilibrium thermodynamics,''
	J. Chem. Phys. \textbf{63} (1975) no.6, 2479
	doi:10.1063/1.431689
	
	\bibitem{Weinhold:1975fyh}
	F.~Weinhold,
	``Metric geometry of equilibrium thermodynamics. II. Scaling, homogeneity, and generalized Gibbs{\textendash}Duhem relations,''
	J. Chem. Phys. \textbf{63} (1975) no.6, 2484
	doi:10.1063/1.431635
	
	\bibitem{Janyszek:1990wdh}
	H.~Janyszek and R.~Mrugaa,
	``Riemannian geometry and stability of ideal quantum gases,''
	J. Phys. A \textbf{23} (1990) no.4, 467
	doi:10.1088/0305-4470/23/4/016
	
	\bibitem{Quevedo:2007mj}
	H.~Quevedo,
	``Geometrothermodynamics of black holes,''
	Gen. Rel. Grav. \textbf{40} (2008), 971-984
	doi:10.1007/s10714-007-0586-0
	[arXiv:0704.3102 [gr-qc]].
	
	\bibitem{Sarkar:2006tg}
	T.~Sarkar, G.~Sengupta and B.~Nath Tiwari,
	``On the thermodynamic geometry of BTZ black holes,''
	JHEP \textbf{11} (2006), 015
	doi:10.1088/1126-6708/2006/11/015
	[arXiv:hep-th/0606084 [hep-th]].
	
	\bibitem{Sahay:2010tx}
	A.~Sahay, T.~Sarkar and G.~Sengupta,
	``On the Thermodynamic Geometry and Critical Phenomena of AdS Black Holes,''
	JHEP \textbf{07} (2010), 082
	doi:10.1007/JHEP07(2010)082
	[arXiv:1004.1625 [hep-th]].
	
	\bibitem{Sarkar:2008ji}
	T.~Sarkar, G.~Sengupta and B.~Nath Tiwari,
	``Thermodynamic Geometry and Extremal Black Holes in String Theory,''
	JHEP \textbf{10} (2008), 076
	doi:10.1088/1126-6708/2008/10/076
	[arXiv:0806.3513 [hep-th]].
	
	
	\bibitem{Bellucci:2008kh}
	S.~Bellucci and B.~Nath Tiwari,
	``On the Microscopic Perspective of Black Branes Thermodynamic Geometry,''
	Entropy \textbf{12} (2010), 2097-2143
	doi:10.3390/e12102097
	[arXiv:0808.3921 [hep-th]].
	
	
	
	
	\bibitem{Kubiznak:2012wp}
	D.~Kubiznak and R.~B.~Mann,
	``P-V criticality of charged AdS black holes,''
	JHEP \textbf{07} (2012), 033
	doi:10.1007/JHEP07(2012)033
	[arXiv:1205.0559 [hep-th]].
	
	
	\bibitem{Kastor:2009wy}
	D.~Kastor, S.~Ray and J.~Traschen,
	``Enthalpy and the Mechanics of AdS Black Holes,''
	Class. Quant. Grav. \textbf{26} (2009), 195011
	doi:10.1088/0264-9381/26/19/195011
	[arXiv:0904.2765 [hep-th]].
	
	\bibitem{Gregory:2017sor}
	R.~Gregory, D.~Kastor and J.~Traschen,
	``Black Hole Thermodynamics with Dynamical Lambda,''
	JHEP \textbf{10} (2017), 118
	doi:10.1007/JHEP10(2017)118
	[arXiv:1707.06586 [hep-th]].
	
	\bibitem{Mann:2025xrb}
	R.~B.~Mann,
	``Black hole chemistry: The first 15 years,''
	Int. J. Mod. Phys. D \textbf{34} (2025) no.09, 2542001
	doi:10.1142/S0218271825420015
	
	\bibitem{AyonBeato:1998ub} E.~Ayon-Beato and A.~Garcia, 
	``Regular black hole in general relativity coupled to nonlinear electrodynamics,'' Phys.\ Rev.\ Lett.\ \textbf{80}, 5056-5059 (1998) 
	doi:10.1103/PhysRevLett.80.5056 
	[arXiv:gr-qc/9911046 [gr-qc]]. 
	
	\bibitem{AyonBeato:1999ec} E.~Ayon-Beato and A.~Garcia, 
	``Nonsingular charged black hole solution for nonlinear source,'' 
	Gen.\ Rel.\ Grav.\ \textbf{31}, 629-633 (1999) 
	doi:10.1023/A:1026640911319 
	[arXiv:gr-qc/9911084 [gr-qc]].
	
	\bibitem{AyonBeato:1999rg} 
	E.~Ayon-Beato and A.~Garcia,
	``New regular black hole solution from nonlinear electrodynamics,'' 
	Phys.\ Lett.\ B \textbf{464}, 25 (1999) 
	doi:10.1016/S0370-2693(99)01038-2 
	[arXiv:hep-th/9911174 [hep-th]]. 
	
	
	\bibitem{Babichev:2020qpr} 
	E.~Babichev, C.~Charmousis, A.~Cisterna and M.~Hassaine,
	``Regular black holes via the Kerr-Schild construction in DHOST theories,''
	arXiv:2004.00597 [hep-th].
	
	\bibitem{Frolov:2016pav} 
	V.~P.~Frolov,
	``Notes on nonsingular models of black holes,''
	Phys.\ Rev.\ D {\bf 94}, no. 10, 104056 (2016)
	doi:10.1103/PhysRevD.94.104056
	[arXiv:1609.01758 [gr-qc]].
	
	
	
	\bibitem{Spallucci:2017aod}
	E.~Spallucci and A.~Smailagic,
	``Regular black holes from semi-classical down to Planckian size,''
	Int.\ J.\ Mod.\ Phys.\ D \textbf{26} (2017) no.07, 1730013
	doi:10.1142/S0218271817300130
	[arXiv:1701.04592 [hep-th]].
	
	
	
	
	
	\bibitem{Feng:2015sbw} 
	X.~H.~Feng and H.~Lu,
	``Higher-Derivative Gravity with Non-minimally Coupled Maxwell Field,''
	Eur.\ Phys.\ J.\ C {\bf 76}, no. 4, 178 (2016)
	 doi:10.1140/epjc/s10052-016-4007-y
	[arXiv:1512.09153 [hep-th]].
	
	
	\bibitem{Cisterna:2020kde} 
	A.~Cisterna, S.~Fuenzalida and J.~Oliva,
	``Lovelock black p-branes with fluxes,''
	Phys.\ Rev.\ D {\bf 101}, no. 6, 064055 (2020)
	doi:10.1103/PhysRevD.101.064055
	[arXiv:2001.00788 [hep-th]].
	
	
	
	\bibitem{Wheeler:1985qd} 
	J.~T.~Wheeler,
	``Symmetric Solutions to the Maximally {Gauss-Bonnet} Extended Einstein Equations,''
	Nucl.\ Phys.\ B {\bf 273}, 732 (1986).
	doi:10.1016/0550-3213(86)90388-3
	
	
	\bibitem{OlivaRay} 
	J.~Oliva and S.~Ray,
	``A new cubic theory of gravity in five dimensions: Black hole, Birkhoff's theorem and C-function,''
	Class.\ Quant.\ Grav.\  {\bf 27}, 225002 (2010)
	 doi:10.1088/0264-9381/27/22/225002
	[arXiv:1003.4773 [gr-qc]].
	
	
	\bibitem{Myers-Robinson} 
	R.~C.~Myers and B.~Robinson,
	``Black Holes in Quasi-topological Gravity,''
	JHEP {\bf 1008}, 067 (2010)
	 doi:10.1007/JHEP08(2010)067
	[arXiv:1003.5357 [gr-qc]].
	
	
	\bibitem{Dehghani-Mann} 
	M.~H.~Dehghani, A.~Bazrafshan, R.~B.~Mann, M.~R.~Mehdizadeh, M.~Ghanaatian and M.~H.~Vahidinia,
	``Black Holes in Quartic Quasitopological Gravity,''
	Phys.\ Rev.\ D {\bf 85}, 104009 (2012)
	 doi:10.1103/PhysRevD.85.104009
	[arXiv:1109.4708 [hep-th]].
	
	
	\bibitem{Quintic} 
	A.~Cisterna, L.~Guajardo, M.~Hassaine and J.~Oliva,
	``Quintic quasi-topological gravity,''
	JHEP {\bf 1704}, 066 (2017)
	 doi:10.1007/JHEP04(2017)066
	[arXiv:1702.04676 [hep-th]].
	
	\bibitem{PablosRecursive} 
	P.~Bueno, P.~A.~Cano and R.~A.~Hennigar,
	``(Generalized) quasi-topological gravities at all orders,''
	Class.\ Quant.\ Grav.\  {\bf 37}, no. 1, 015002 (2020)
	 doi:10.1088/1361-6382/ab5410
	[arXiv:1909.07983 [hep-th]].
	
	
	\bibitem{Ferraro}
	M.~Aiello, R.~Ferraro and G.~Giribet,
	``Exact solutions of Lovelock-Born-Infeld black holes,''
	Phys.\ Rev.\ D \textbf{70}, 104014 (2004)
	doi:10.1103/PhysRevD.70.104014
	[arXiv:gr-qc/0408078 [gr-qc]].
	
	\bibitem{Banados:1992wn} 
	M.~Banados, C.~Teitelboim and J.~Zanelli,
	``The Black hole in three-dimensional space-time,''
	Phys.\ Rev.\ Lett.\  {\bf 69}, 1849 (1992)
	doi:10.1103/PhysRevLett.69.1849
	[hep-th/9204099].
	
	
	
	\bibitem{Wald:1999vt}
	R.~M.~Wald,
	``The thermodynamics of black holes,''
	Living Rev. Rel. \textbf{4} (2001), 6
	doi:10.12942/lrr-2001-6
	[arXiv:gr-qc/9912119 [gr-qc]].
	
	\bibitem{Bekenstein:1973ur}
	J.~D.~Bekenstein,
	``Black holes and entropy,''
	Phys. Rev. D \textbf{7} (1973), 2333-2346
	doi:10.1103/PhysRevD.7.2333
	
	\bibitem{Bekenstein:1974ax}
	J.~D.~Bekenstein,
	``Generalized second law of thermodynamics in black hole physics,''
	Phys. Rev. D \textbf{9} (1974), 3292-3300
	doi:10.1103/PhysRevD.9.3292
	
	\bibitem{Bardeen:1973gs}
	J.~M.~Bardeen, B.~Carter and S.~W.~Hawking,
	``The Four laws of black hole mechanics,''
	Commun. Math. Phys. \textbf{31} (1973), 161-170
	doi:10.1007/BF01645742
	
	\bibitem{Hawking:1975vcx}
	S.~W.~Hawking,
	``Particle Creation by Black Holes,''
	Commun. Math. Phys. \textbf{43} (1975), 199-220
	doi:10.1007/BF02345020
	
	
	
	
	
	\bibitem{Wei:2019yvs}
	S.~W.~Wei, Y.~X.~Liu and R.~B.~Mann,
	``Ruppeiner Geometry, Phase Transitions, and the Microstructure of Charged AdS Black Holes,''
	Phys. Rev. D \textbf{100} (2019) no.12, 124033
	doi:10.1103/PhysRevD.100.124033
	[arXiv:1909.03887 [gr-qc]].
	
	\bibitem{Wei:2015iwa}
	S.~W.~Wei and Y.~X.~Liu,
	``Insight into the Microscopic Structure of an AdS Black Hole from a Thermodynamical Phase Transition,''
	Phys. Rev. Lett. \textbf{115} (2015) no.11, 111302
	doi:10.1103/PhysRevLett.115.111302
	[arXiv:1502.00386 [gr-qc]].
	
	\bibitem{Ghosh:2019rsu}
	A.~Ghosh and C.~Bhamidipati,
	``Contact Geometry and Thermodynamics of Black Holes in AdS Spacetimes,''
	Phys. Rev. D \textbf{100} (2019) no.12, 126020
	doi:10.1103/PhysRevD.100.126020
	[arXiv:1909.11506 [hep-th]].
	
	\bibitem{Ghosh:2019pwy}
	A.~Ghosh and C.~Bhamidipati,
	``Thermodynamic geometry for charged Gauss-Bonnet black holes in AdS spacetimes,''
	Phys. Rev. D \textbf{101} (2020) no.4, 046005
	doi:10.1103/PhysRevD.101.046005
	[arXiv:1911.06280 [gr-qc]].
	
	\bibitem{Yerra:2021hnh}
	P.~K.~Yerra and C.~Bhamidipati,
	``Novel relations in massive gravity at Hawking-Page transition,''
	Phys. Rev. D \textbf{104} (2021) no.10, 104049
	doi:10.1103/PhysRevD.104.104049
	[arXiv:2107.04504 [gr-qc]].
	
	\bibitem{Singh:2023ufh}
	A.~Singh,
	``Thermodynamic geometry of dyonic black holes in AdS in extended phase space,''
	Mod. Phys. Lett. A \textbf{38} (2023) no.40, 2350173
	doi:10.1142/S0217732323501730
	
	\bibitem{Singh:2025ueu}
	A.~Singh and S.~Mahish,
	``Criticality of charged AdS black holes with string clouds in boundary conformal field theory,''
	[arXiv:2504.20486 [hep-th]].
	
	\bibitem{Singh:2023hit}
	A.~Singh, P.~Mukherjee and C.~Bhamidipati,
	``Thermodynamic curvature of charged black holes with AdS2 horizons,''
	Phys. Rev. D \textbf{108} (2023) no.10, 106011
	doi:10.1103/PhysRevD.108.106011
	[arXiv:2307.11641 [hep-th]].
	
	\bibitem{Xu:2020gud}
	Z.~M.~Xu, B.~Wu and W.~L.~Yang,
	``Ruppeiner thermodynamic geometry for the Schwarzschild-AdS black hole,''
	Phys. Rev. D \textbf{101} (2020) no.2, 024018
	doi:10.1103/PhysRevD.101.024018
	[arXiv:1910.12182 [gr-qc]].
	
	\bibitem{May:2013yxc}
	H.~O.~May, P.~Mausbach and G.~Ruppeiner,
	``Thermodynamic curvature for attractive and repulsive intermolecular forces,''
	Phys. Rev. E \textbf{88} (2013) no.3, 032123
	doi:10.1103/PhysRevE.88.032123
	
	\bibitem{Zhang:2019neb}
	B.~Zhang, S.~S.~Wan and M.~Ruggieri,
	``Thermodynamic Geometry of the Quark-Meson Model,''
	Phys. Rev. D \textbf{101} (2020) no.1, 016014
	doi:10.1103/PhysRevD.101.016014
	[arXiv:1907.11781 [hep-ph]].
	
	\bibitem{Anand:2025mlc}
	A.~Anand, S.~Noori Gashti and A.~Singh,
	``Thermodynamic curvature and topological insights of Hayward black holes with string fluids,''
	Phys. Dark Univ. \textbf{49} (2025), 101994
	doi:10.1016/j.dark.2025.101994
	[arXiv:2506.23736 [gr-qc]].
	
	
	
	
	\bibitem{Singh:2020tkf}
	A.~Singh, A.~Ghosh and C.~Bhamidipati,
	``Thermodynamic curvature of AdS black holes with dark energy,''
	Front. in Phys. \textbf{9} (2021), 65
	doi:10.3389/fphy.2021.631471
	[arXiv:2002.08787 [gr-qc]].
	
	\bibitem{Mahish:2020gwg}
	S.~Mahish, A.~Ghosh and C.~Bhamidipati,
	``Thermodynamic curvature of the Schwarzschild-AdS black hole and Bose condensation,''
	Phys. Lett. B \textbf{811} (2020), 135958
	doi:10.1016/j.physletb.2020.135958
	[arXiv:2006.02943 [hep-th]].
	

	
	
	
	\bibitem{Yerra:2020oph}
	P.~K.~Yerra and C.~Bhamidipati,
	``Ruppeiner Geometry, Phase Transitions and Microstructures of Black Holes in Massive Gravity,''
	Int. J. Mod. Phys. A \textbf{35} (2020) no.22, 2050120
	doi:10.1142/S0217751X20501201
	[arXiv:2006.07775 [hep-th]].
	
	\bibitem{Wu:2020fij}
	B.~Wu, C.~Wang, Z.~M.~Xu and W.~L.~Yang,
	``Ruppeiner geometry and thermodynamic phase transition of the black hole in massive gravity,''
	Eur. Phys. J. C \textbf{81} (2021) no.7, 626
	doi:10.1140/epjc/s10052-021-09407-y
	[arXiv:2006.09021 [gr-qc]].
	
	\bibitem{NaveenaKumara:2020biu}
	A.~Naveena Kumara, C.~L.~Ahmed Rizwan, K.~Hegde, M.~S.~Ali and K.~M.~Ajith,
	``Ruppeiner geometry, reentrant phase transition, and microstructure of Born-Infeld AdS black hole,''
	Phys. Rev. D \textbf{103} (2021) no.4, 044025
	doi:10.1103/PhysRevD.103.044025
	[arXiv:2007.07861 [gr-qc]].
	

	
	
	
	
	
	
	
	
	
	
	
	
\end{thebibliography}
\end{document}